\documentstyle[12pt,aasms4]{article}

\def\gs{\mathrel{\raise0.35ex\hbox{$\scriptstyle >$}\kern-0.6em 
\lower0.40ex\hbox{{$\scriptstyle \sim$}}}}
\def\ls{\mathrel{\raise0.35ex\hbox{$\scriptstyle <$}\kern-0.6em 
\lower0.40ex\hbox{{$\scriptstyle \sim$}}}}

\begin{document}

\title{Constraints on the Evolution of S0 Galaxies in Rich Clusters at 
Moderate Redshift\footnotemark}

\author{Lewis Jones,$^2$ Ian Smail$^3$ \& Warrick J.\ Couch$^2$} 
\affil{\tiny 2) School of Physics, University of New South Wales, Sydney 2052, Australia}
\affil{\tiny 3) Department of Physics, University of Durham, South Road, Durham, DH1 3LE, UK}

\footnotetext{Based on observations with the NASA/ESA {\it Hubble
Space Telescope} obtained at the Space Telescope Science Institute,
which is operated by the Association of Universities for Research in
Astronomy Inc., under NASA contract NAS 5-26555.} 

\begin{abstract}
We combine morphological classifications from deep {\it Hubble Space
Telescope} ({\it HST}\,) imaging of a sample of three clusters at
$z=0.31$ and a further nine clusters at $z=0.37$--0.56 with existing
spectroscopic observations of their elliptical (E) and S0 populations, to
study the relative spectral properties of these two galaxy types.   We have
also used spectroscopic and imaging data in the Coma cluster as a
present day example of a rich cluster environment with which to compare
our data at higher redshift.  These data span the range where strong 
evolution is claimed in the proportion of S0 galaxies within rich 
clusters.  Techniques have been recently developed
to analyse the strengths of absorption lines in the spectra of local,
passive galaxies to separate the effects of age and metallicity in the
spectra and hence date the ages of the most recent, substantial star
formation episode in these galaxies. We show that the spectroscopic
data in the distant clusters is of sufficient quality to allow us to
apply these techniques and use them to determine the relative ages for the E
and S0 populations in these distant clusters.  We then compare the
`ages' for each type to search for the signature
of the recent formation of the bulk of the S0 population as suggested
by Dressler et al.\ (1997).  We find {\em no} statistically significant
difference between the luminosity weighted ages of the E and S0
galaxies in these clusters.  We translate this into a limit such that
no more than half of the galaxies in the clusters at $z=0.31$ have
undergone a burst of star formation ($>11$\% by mass) in the 1\,Gyr
prior to the observations.  Our results, in conjunction with other
work, suggests that the progenitors of the S0 galaxies in rich clusters
are mostly early type spirals who, through interactions with the
cluster environment, have had their star formation truncated.  This indicates
a relatively unspectacular origin for the missing S0 population.
\end{abstract}


\section{Introduction}

The conventional picture of luminous early type cluster galaxies (E and
S0) is that their stars formed in a short period of time nearly a
Hubble time ago (Tinsley \& Gunn 1976).  Since the suggestion of
Sandage \& Visvinathan (1978) that the UV-optical color-magnitude
relation for cluster early type galaxies could be used to constrain
their star formation histories, studies of the cluster color-magnitude
relations have almost universally confirmed the conventional picture
(Bower, Lucey, \& Ellis 1992, BLE; Ellis et al.\ 1997); the luminous,
early type populations of rich clusters are remarkably homogeneous and
the bulk of their stellar populations appear to be old.  

Dressler et al.\ (1997, D97), on the other hand, suggest that the
luminous S0 galaxies are less abundant 
in distant clusters relative to the fractions of these types seen in present
day clusters.  They suggest that there must have been strong
evolution in the S0 population across the redshift range $z=0$--0.5 to
account for the large S0 populations seen in local clusters.  If S0
galaxies are being produced in clusters in recent times then they
should appear characteristically younger than the elliptical
population.  Indeed, Kuntschner \& Davies (1998), find such an effect
with the majority of the S0 galaxies in a sample drawn from the Fornax
cluster having luminosity-weighted ages several Gyrs younger than the
ellipticals.   Moreover, the post-starburst galaxies (PSG) found in
large numbers in the distant clusters (e.g.\ Couch \& Sharples 1987,
CS87; Dressler et al.\ 1999, D99) would provide an obvious intermediate
phase in the transformation of star forming disk galaxies into passive
S0's (Poggianti et al.\ 1999, P99).  However, this model then leads to
a potential conflict with those results confirming the conventional
model of old, coeval E and S0 galaxies.

In this Letter, we apply the spectroscopic technique of Kuntschner \&
Davies (1998) in clusters out to $z\sim 0.5$ to place new constraints
on the evolution of the S0 population in rich clusters by a direct
comparison of the relative ages of the cluster S0 and E populations at
three epochs spanning the range where strong evolution is claimed in
the proportion of S0 galaxies within rich clusters (D97).  In \S2, we
discuss the morphological and spectroscopic data as well as describing
the method we used to separate age and metallicity effects in the
spectra, then in \S3, we present the derived constraints on S0 evolution,
and finally in \S4, we suggest a resolution to the apparent conflict.

\section{Observations and Analysis}

Our analysis makes use of existing moderate resolution spectroscopy and
{\it HST} morphologies for galaxies in three clusters at $z=0.31$ and a
further 9 clusters between $z=0.37$--0.55. Existing spectroscopic and
morphological data for Coma are included as a local ($z\sim$0.0), rich
cluster comparison.  In particular we highlight
the usefulness of three clusters at $z=0.31$ from our sample in this
analysis -- for the evolution claimed by D97, almost half the S0
population of these clusters should have been produced in the 1.5\,Gyrs
since $z=0.5$.  This relatively high fraction coupled with the
reasonable numbers of S0 galaxies in these systems provides a strong
test of the D97 scenario. 

\subsection{Morphological Samples}

For Coma, the morphologies are taken from Godwin, Metcalfe, \& Peach
(1983).  There are 63 elliptical and 121 S0 cluster members for which
we also have high quality spectra.

For the three clusters at $z=0.31$ (AC\,103, AC\,114, and AC\,118), we
have used the morphological classifications from Couch et al.\ (1998, C98)
and the details of the observations are contained therein.  The F702W
imaging consists of 3-orbit exposures of the cores of AC\,118 and
AC\,103, and a mosaic of four, 6 orbit exposures on AC\,114.  The
morphologies from Couch et al.\ (1998) are robust down to
$R_{702}\sim22.5$ yielding classifications for 25 spectroscopically-confirmed cluster ellipticals and 29 S0 cluster members.

Our more distant cluster sample comes from observations by the MORPHS
collaboration, consisting of deep {\it HST} imaging in the F702W, F814W, or
F555W
filters of 11 fields in 10 clusters in the redshift range $0.37 \leq z
\leq 0.56$.  Details of these observations can be found in Smail et
al.\ (1997).  The spectroscopic follow-up of this survey provides
membership for 15 elliptical and 10 S0 galaxies brighter than $R$=23.5
or $I$=23.0 (D99).  Smail et al.\ (1997) and D97 have
shown that the incidence of round and/or face-on S0 galaxies likely to be
misclassified as ellipticals in the distant clusters 
is no worse than that seen in local samples (15\%).

\subsection{Spectroscopic Observations}

The Coma cluster data have been published by Caldwell et al.\ (1993)
and Caldwell \& Rose (1997, CR97), and the reduced galaxy spectra were kindly
provided by the authors.  The spectral resolution of their spectra is
$\sim4$\,\AA\ FWHM with S/N ranging from 30 to 100 -- substantially
above that achieved for individual galaxies in the distant samples.

Spectroscopic observations for the three clusters at $z=0.31$ are taken
from two sources, CS87 and C98.  The CS87 spectra were
collected at the 3.9-m Anglo-Australian Telescope (AAT) with the FOCAP
multi-fiber system and the IPCS detector, and have a spectral
resolution of $\sim 4$\,\AA\ FWHM.  The C98 spectra
were also collected at the AAT with LDSS-1, a multi-slit spectrograph, and
have a spectral resolution of $\sim 9$\,\AA\  FWHM.  The spectra in
both samples range in signal-to-noise ratio (S/N) from 3 to 10, with
the median being $\sim$8.

The spectroscopic observations of galaxies in the MORPHS sample is
presented in D99 and these spectra were kindly
provided by the authors.  The majority of these spectra come from the
COSMIC multi-object spectrograph on the 5.1-m Hale Telescope.  The
spectral resolution of these observations is 8\,\AA\ FWHM and the
median S/N is 25 for the spectra analysed here (the lowest having
S/N=5).

The majority of the spectra provide rest-frame wavelength coverage which includes 
3900\,$\leq\lambda\leq$\,5300\,\AA, i.e.\ from the Ca{\sc ii} H \& K
lines through to the Mg triplet.  All of the galaxy spectra have been
smoothed to the 9 \AA\ resolution of the LDSS-1 data.

\subsection{Spectral Indices}

The main obstacle to an analysis of ages and metallicities of galaxies
is the degeneracy of those two effects on the galaxy spectral energy
distribution.  Focusing on individual spectral features has allowed
marked progress in breaking that degeneracy (c.f.\ Worthey 1994, W94;
Worthey \& Ottaviani 1997; Jones \& Worthey 1995), and for this paper
we have chosen seven of the commonly used Lick spectral indices.  The
H$\delta_{A}$ index acts as an age indicator (Worthey \& Ottaviani 1997), 
while the Ca4227, Fe4383, Ca4455, Fe4531, C$_{2}$4668, and Fe5015 indices
trace metallicity effects (W94). This analysis relies upon high
quality spectra to robustly separate the effects of age and metallicity,
and given that we are looking for evidence of evolution in the characteristics
of the overall population of S0 galaxies {\it relative} to ellipticals, 
we have chosen to co-add the S0 and elliptical spectra at each redshift 
and perform a {\it differential} (E versus S0) analysis in each case 
(z=0.0, 0.31, $\sim$0.50). Co-addition of the spectra was achieved using 
the SCOMBINE routine in IRAF after having first transformed them onto 
the rest wavelength scale. This provided final spectra with S/N$=40$ and
43 for the E and S0's in the $z=0.31$ clusters, and S/N$=97$ and 79 for
the same respective types in the $z\sim 0.5$ clusters. Seven indices were 
measured in the co-added spectra, with those pertaining to the same
element being combined to form average values. 

Figure 1 shall be used to present our results. The grid lines 
represent lines of constant age (dotted lines) and constant metallicity 
(solid lines) as taken from the single stellar population models of 
W94; they are labelled in Panel c to indicate age differences in Gyr
and metallicity differences in dex units of [Fe/H]. Each redshift is 
represented by a different shaped symbol with open and filled versions 
being used to differentiate between the E and S0 measurements, respectively. 
Regarding the separation of age and metallicity, we see that while the 
age and metallicity lines are not orthogonal, they do indeed provide  
discrimination between these two effects.  Note that any galaxy `ages' 
discussed in this Letter will be luminosity-weighted mean ages, and are 
not meant to indicate the ages of the oldest stars in the galaxies.  
Also, because the three different samples of spectra have been 
collected with different telescope--instrument--detector combinations, 
and because we have not attempted to reconcile any offsets between 
spectral index measurements of the different systems and the models, we 
stress that there are zero point offsets relative to the age--metallicity 
model grid associated with each redshift.  Accordingly, we cannot simply 
compare `ages' between redshifts; only the relative ages are meaningful at 
a given redshift.  This limitation does not bear on our analysis since we are
concerned only with a differential comparison between the E and S0 populations
at a single redshift. The following sections address the results derived 
from Fig.~1, and their implications.

\section{Results and Discussion}


\subsection{Main Result}

Panels a--c of Fig.~1 show the H$\delta_{A}$ index plotted against the 
mean of the three Fe indices, the mean of the two Ca indices, and the 
C$_{2}4668$ index, respectively. The C$_{2}4668$ has been proposed as a 
possible metallicity indicator and below we discuss its usefulness in the 
context of our results. The error bars for H$\delta_{A}$ are approximately 
the size of the points for the two higher redshift samples and one fifth 
the size of the points for Coma, so the separation between points in 
H$\delta_{A}$ is less than 1$\sigma$ for the E and S0 spectra at a given 
redshift. Furthermore, in Panels~a \& b, no pair of points is separated by 
more than 1$\sigma$ in the metallicity index. In contrast, we see for
the C$_{2}4668$ index good agreement between the E and S0 populations in 
Coma, some consistency between the populations in the MORPHS sample (less 
than 2$\sigma$ difference), but a 7$\sigma$ difference between the 
apparent metallicities of the E and S0 population in the $z=0.31$ 
clusters from CS87/C98. We note that the CS87/C98 elliptical and S0 
spectra were obtained in the same observing runs and reduced without 
differentiating between the morphological types, so it is difficult to 
imagine that a systematic error in the reduction could be responsible for 
this disagreement. The near orthogonality of the age and metallicity
sequences with the index axes means that the  difference in the C$_{2}4668$ 
index for elliptical and S0 galaxies in the CS87/C98 sample translates 
into an age difference of around 1--2 Gyrs.

The discrepent results from the C$_{2}4668$ index compared to the broad
agreement between our other two age/metallicity indicators requires an
explanation.  The C$_{2}4668$ index is extremely sensitive to changes
in the photospheric carbon abundance in stars (Tripicco \& Bell 1995)
and the most striking change in the carbon abundance of a star is in
the asymptotic giant branch (AGB) phase of evolution where stars reach
the thermally pulsing AGB and become Carbon stars (Iben 1993).  The
brightest {\em and} coolest\footnote{They must be cool
(K and M) stars because they would otherwise not have formed molecular
carbon.} of these stars have high initial masses and will be extinct by
about 1 Gyr after a burst of star formation suggesting that a recent
burst is indeed  the cause of the difference between the E and S0
C$_{2}4668$ indices. However, while the integrated number of Carbon
stars produced over the first 1 Gyr after a burst is enough to produce
the observed change in the C$_{2}4668$ index, the number of such stars
present {\em at any one time} is too small (Buzzoni 1989).  Moreover,
the similarity between the E and S0 galaxies at other redshifts (unless
the three CS87/C98 clusters are conspiratorily pathological), suggests that
we cannot appeal to global differences in carbon abundances due to
environment.  The lack of a simple explanation of the discrepent
C$_{2}4668$ indices leads us to concentrate on our other
age/metallicity indicators and we suggest that the C$_{2}4668$ index,
despite its empirical connection with metallicity demonstrated by
W94, requires more study before it can be reliably used to
investigate the age/metallicity of integrated stellar populations.   

In summary, with the exception of the CS87/C98 C$_{2}4668$ index in the $z=0.31$ clusters, we find that
the luminosity weighted average age of elliptical and S0 galaxies in
rich clusters derived from this data are indistinguishable out to
$z\sim0.5$.

\subsection{Star Formation Histories}

The distribution of H$\delta_{A}$ values for E and S0 galaxies in the
different clusters can also be used to place limits on recent star
formation in the galaxies, and hence we have measured this spectral
index in the individual galaxy spectra as well.  Comparing
the E and S0 H$\delta_{A}$ distributions for the three different cluster
samples, we find using the KS test that they are indistinguishable
with the formal probabilities that they are drawn from different 
populations being: P(KS)$_{\rm Coma}$=53\%, P(KS)$_{\rm CS87/C98}$=34\%, 
and P(KS)$_{\rm MORPHS}$=0\%.  

The D97 results suggest that the number of S0 galaxies
in the more distant MORPHS clusters would have to increase four-fold to
reproduce the E:S0 ratio in local clusters, while 50\% of the S0's in
the $z=0.31$ samples would need to have formed since $z\sim0.5$.  We
can directly evaluate the sensitivity of our analysis to this form of
evolution by selecting a random sample of 50\% of the $z=0.31$ S0's and
adding a contribution from a model spectrum of a young stellar
population from the library of Bruzual \& Charlot (1993) to each S0
spectrum and then recomputing the spectral indices.  We have done this
with two solar metallicity models, taken 0.5 and 1.0\,Gyrs after the end
of a  burst of star formation.  We then use the new indices to
recalculate the H$\delta_{A}$ distribution of the entire S0 sample, and compare
the new S0+model H$\delta_{A}$ distribution to the original E
H$\delta_{A}$ distribution.  This allows us to determine the minimum
contribution from a young population in the S0 sample which our
analysis of the $z=0.31$ clusters would detect (assuming the broad
framework of the evolution claimed by D97). The resulting constraints,
for the two different models, are that the H$\delta_{A}$ distribution
for the S0+model galaxies is statistically distinguishable from the
ellipticals at $\ge 96$\% (92\%) significance level when 10\% 
(10\%) of the $V$-band light in half of the S0 sample is contributed by a
stars from a 0.5 (1.0)\,Gyr old post-starburst stellar
population.\footnote{This corresponds to a change in the average S0
H$\delta_{A}$ index of $\sim$0.5\AA, which, given that only 50\% of the
S0 population has the  young star light added, means the average index
change we are sensitive to in the post-starburst S0 galaxies is
$\sim$1.0\AA.}

Converting this constraint to the mass fraction of a galaxy converted to stars
in the burst, we conclude that the S0 galaxies at $z=0.31$ have not
experienced bursts of more than 6\% (11\%) by mass in the 0.5\,Gyr
(1\,Gyr) prior to the time of observation (Bruzual \& Charlot 1993).  
These modest burst strengths are consistent with the
strength of the star bursts found by Barger et al.\ (1996) in these
same clusters.  The equivalent star formation rates are typical of
those seen in normal spiral galaxies.  We conclude that for our analysis 
to detect a change in the characteristic age of S0 galaxies relative to 
ellipticals at the same redshift, at least half of the S0's in these 
clusters would have to have undergone particularly intense bursts of star 
formation ($>11$\% of their current stellar mass) in the 1\,Gyr prior to 
observation, i.e.\ between $z=0.3$ and $z\sim 0.5$.  {\it This limit 
suggests that the bulk of the cluster S0 population is not being built up 
through a process of intense starbursts where spirals are converting 
their entire gas reservoir into stars in a single event.}

\section{Conclusions}

By combining high resolution {\it HST} imagery and existing
spectroscopy, we have been able to investigate the evolution of early
type galaxies in moderate redshift clusters.  We have used the stellar
population models of Worthey (1994) to disentangle the effects
of age and metallicity in our spectra. We have two main conclusions: Firstly,
we find no evidence for an age difference between E and S0 galaxies in rich
clusters out to $z\sim$0.5. Secondly, the precision with which our 
spectral index method rules out an age difference would imply that any 
recent star formation in the cluster S0's at $z=0.31$ could have involved 
no more than 50\% of these galaxies undergoing bursts in which no more
than $\sim 11$\% of their mass was converted into stars.   

Poggianti et al.\ (1999) have analysed the spectroscopic and
morphological catalogs gathered for the $z=0.37$--0.57 clusters by
D99.  They find evidence for a wide-spread
suppression of star formation in cluster members, leading to large
populations of PSG and passive spirals which are much more prevelant in
the cluster environment. The progenitors of these galaxies are
identified with dusty, star
burst galaxies infalling from the surrounding field.  They suggest
that the resulting passive, disk population are the progenitors of the S0
galaxies.  At lower redshifts, Caldwell et al.\ (1996, C96) have shown
that in the Coma cluster, Balmer line strengths of PSG early type
galaxies cannot be explained by a previously constant star formation
rate, typical of spiral galaxies, which has been recently truncated,
but that they require truncation to occur after a brief burst of
$\sim$10\% by mass, and well within the above limits placed on
starbursts at higher redshifts.  In CR97 they conclude that these
final, modest bursts of star formation in Coma, as well as other nearby
rich clusters, are indeed associated with the cluster environment.
Also, in both C96 and CR97, they show that most of the starburst and
post-starburst galaxies in their clusters are, morphologically, early
type disk galaxies, either early type spirals or S0's.

In light of the evidence from D97 that the S0 population in rich
clusters needs to strongly increase between $z\sim 0.5$ and today, the
results of P99, C96 and CR97, together with the results of this Letter,
suggest a mostly unspectacular origin for the missing S0 galaxies.  We
conclude from the assembled evidence that the missing S0 galaxies are
the remnants of old, early-
type spiral galaxies which, as a result of
their interaction with the rich cluster environment, have had their
star formation truncated after possibly undergoing a final, modest
burst of star formation.  This conclusion suggests that with more
sensitive age indicators and spectra of higher signal-to-noise and
higher spectral resolution, we would be able to detect the effects of
those bursts and test this hypothesis.  Such observations are possible
from current 4- and 8-meter class telescopes.

\acknowledgements

The authors would like to thank the MORPHS collaboration for the use of their
spectra in advance of publication, and Jim Rose and Nelson Caldwell for
provding the Coma spectra.  This work was supported by the Australian Research 
Council.  IRS acknowledges additional support from PPARC and a travel grant
from the Royal Society.


\clearpage

\figcaption[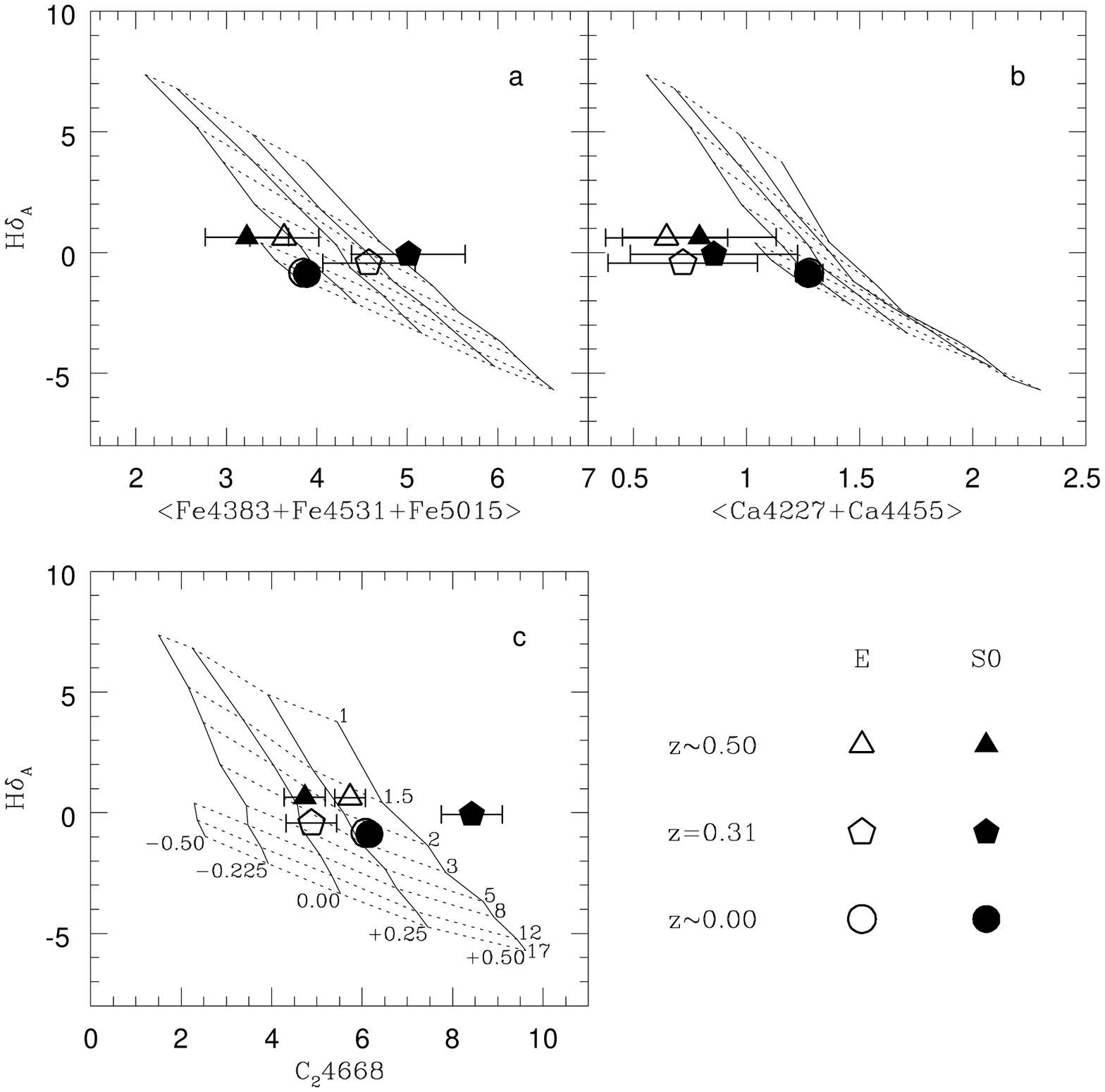]{\label{fig1}
H$\delta_{A}$ index strength plotted against the $<\!{\rm Fe}\!>$ (Panel~a), 
$<\! {\rm Ca}\! >$  (Panel~b), and C$_{2}$4668  (Panel~c) indices, for 
the E and S0 co-added spectra each at $z\sim0.0$, $z=0.31$ and $z\sim0.5$ 
(see legend). The {\it dashed} and {\it solid} grid lines are, 
respectively, the age and metallicity sequences from the models of 
Worthey (1994); those in Panel~c are labelled in units of Gyrs and dex 
[Fe/H].}


\end{document}